\documentstyle[12pt,aps,preprint]{revtex}

\begin{document}
\title{Ideally Efficient Irreversible Molecular Gears}
\author{I.M.Sokolov}
\address{Theoretische Polymerphysik, Universit\"{a}t Freiburg, Hermann-Herder-Str. 3,%
\\
D-79104 Freiburg i.Br., Germany}
\maketitle

\begin{abstract}

Typical man-made locomotive devices use reversible gears, as cranks, for
transforming reciprocating motion into directed one. Such gears are
holonomic and have the transduction efficiency of unity. On the other hand,
a typical gear of molecular motors is a ratchet rectifier, which is
irreversible. We discuss what properties of rectifier mostly influence the
transduction efficiency and show that an apliance which locks under
backwards force can achieve the energetic efficiency of unity, without
approaching reversibility. A prototype device based on ratchet principle is
discussed.

\bigskip

\textbf{PACS No: } 05.40.-a; 05.70.Ln; 87.10.+e

\end{abstract}

\makeatletter
\global\@specialpagefalse

\newpage

Man-made engines powering our cars, trains and ships, and molecular motors
powering cells and subcell units are energy transducer designed to transform
chemical energy, stored in form of fuel and oxygen, into mechanical work.
Both can be considered as consisting of the working unit(s) and of a gear. A
gear is used in order to transform the oscillatory motion of a piston (or a
kinesin molecule) $x(t)$ into a continuous directed motion $X(t)\sim vt$, or
into continuous rotation $\varphi (t)\sim \omega t$. A typical gear used for
technical applications is a crank-and-shaft mechanism. This gear is
reversible, since the continuous rotation of the crank's axle causes the
oscillatory piston's motion. The relation between $x$ and $\varphi $
corresponds to a periodic, locally invertible function. The transformation
of oscillations into directed motion implies symmetry breaking, determining
the direction of motion. Cranks use spontaneous symmetry breaking: here both
rotation directions are possible; the actual one is determined by initial
conditions. The onset of motion is hard: too small oscillations can not be
transformed into a continuous rotation. Moreover, the holonomic nature of
gearing transformation implies synchronization of working units, if several
of them are used. Molecular motors, on the other hand, use rectifiers (such
as a ratchet-and-pawl system) in which the spatial symmetry is lacking from
the very beginning. Rectifiers are irreversible gears, as clearly
illustrated by usual electric appliance: A diode rectifier transforms an
alternating current into direct one, but, being fed with a direct current,
it does not produce an alternating one, but only heat. Rectification has
significant advantages compared to holonomic gearing. Thus, the soft onset
of the motion allows for easy control at small velocities, and the
asynchronous mode of operation is of great virtue in nanoscale cellular
systems, since the synchronization of molecular-level reaction events
(having stochastic, Poissonian character) is a problematic task. This
property is often referred to as the ability to rectify noise \cite
{Magnasco1,Magnasco2,A+M+P}.

The quality of a gear can be characterized by its energetic efficiency, i.e.
by a quotient between the input energy and useful work performed, so that
the question of energetics of gearing got recently much attention \cite
{Bierast,Sekimoto,Sok+Blum,Takagi1,Takagi2,Parmeggiani} within 
different theoretical frameworks. The
energetic efficiency of a holonomic gear is unity, and the Second law of
thermodynamics implies that the energetic efficiency of any other isothermal
gear can not exceed this limit. On the other hand, typical efficiencies of
prototype ratchets (as a rocked system of Refs. \cite{Takagi1,Takagi2}
transporting particles against constant outer potential due to the work of
additive oscillating field) in quasistatic regime (guaranteeing the no-loss
condition for typical thermodynamic appliances, Ref.\cite{Bailyn}) are so
poor, that one wonders, why didn't the Nature look for anther mechanism to
do the work. The exception is one of the systems (''system b'')) discussed in
Ref.\cite{Parmeggiani} which is essentially a synchronized, quasiequilibrium
motor. In what follows we analyze in detail the thermodynamics of
rectification and show that prototypical ratchet devices lack an important
property of effective rectifiers, namely the backward locking (known from
the common experience with the macroscopic ratchet-and-pawl mechanism). As
we proceed to show, an ideal rectifier can perform as good as a crank, and
moreover a minor variation of a simple ratchet rectifier can produce a gear
whose performance is not too far from the ideal one.

Let us discuss the work produced by an isothermal system under changes of
outer conditions. The mean energy of the system is given by $%
E=\sum_{i}e_{i}p_{i}$, where $e_{i}$ is the energy of a (micro)state $i$ and 
$p_{i}$ is a corresponding probability (occupation number). The energy
change is then given by 
\begin{equation}
dE=\sum_{i}de_{i}p_{i}+\sum_{i}e_{i}dp_{i}.  \label{dE}
\end{equation}
In quasiequilibrium Eq.(\ref{dE}) corresponds to the form $dE=\delta
A+\delta Q$ of the First Law of thermodynamics. Out of equilibrium, the
first term still corresponds to the work of outer forces, but the second one
shows some new, typically nonequilibrium, aspects.

Let us discuss a case when $i$ can be parametrized by continuous phase space
coordinates $\mathbf{r=(x,\dot{x})}$. In an overdamped situation (typical
for biological systems) the kinetic degrees of freedom decouple from spatial
ones, $p({\mathbf{r}})=p({\mathbf{x}})p({\mathbf{\dot{x}}})$, with$\ p({\mathbf{%
\dot{x})}}$ being equilibrium Maxwell distribution,\ see Ref. \cite{Risken}.
Thus we can fully concentrate on the coordinate space of the system. The
energy changes due to the redistribution of occupation probabilities during
time $dt$ can be expressed as: 
\begin{equation}
\sum_{i}e_{i}dp_{i}=dt\int_{V}e({\mathbf{r}})\frac{dp({\mathbf{r}})}{dt}d{\mathbf{r}}=%
dt\left[ \int_{V}{\mathbf{j(x)}}grad\,U({\mathbf{x}})dV-\int_{dV}U({\mathbf{x}})%
{\mathbf{j}}({\mathbf{x}})d{\mathbf{\sigma }}\right]
\end{equation}
where the continuity equation $dp({\mathbf{x}})/dt+div{\mathbf{j(x)}}=0$ in
coordinate space is used. Here $d\mathbf{\sigma }$ denotes the surface
element of the system's outer boundary. The first term represents the heat
absorbed from the bath per unit time and is equal to the Joule heat taken
with an opposite sign. The second, surface term describes the work (produced
within the system's volume) of the currents, which are generated outside of
the system. The energy balance in the system reads: $dE/dt=P_{F}+P_{I}+q,$
where $P_{F}$ is the power of outer forces, $P_{I}$ is the power of outer
currents, and $q$ is the heat absorbed by a system from the heat bath per
unit time. For a device  acting periodically or under stochastic force with zero mean $%
\overline{dE/dt}=0$, so that $\overline{P_{F}}+\overline{P_{I}}+\overline{q}%
=0$. Depending on the particular arrangement, the input work and the useful
work of a gear can be differently distributed between $P_{F}$ and $P_{I}$.
On the other hand the mean heat is always dissipated, $\overline{q}<0.$

As an example let us consider a typical electrical arrangement consisting of
an outer a.c. source of voltage $U_{F}(t)$, of a rectifier, and of an
accumulator (maintaining a constant voltage $\Delta U$) switched in series,
see the insert in Fig.1. If a thermodynamic appliance achieves ideal
efficiency, it typically achieves it in quasistatic regime, since a
finite-velocity mode of operation is inevitably connected with losses, Ref.%
\cite{Bailyn}. Confining ourselves to a quasistatic situation, we can
describe a rectifier by a Volt-Amper characteristics (load-current
characteristic, LCC) $I(t)=I(U(t))$: The state of the whole system is
characterized by the current $I(t)$ being the function of $U$, the potential
difference at the rectifier. The useful work (charging the battery) is
produced by the outer currents flowing against the batterie's voltage, so
that its value per unit time is $P=-P_{I}=-\Delta UI(t)$, and the Joule heat $%
Q=-q=U(t)I(t)$ is uselessly dissipated. The energy
balance discussed before corresponds to a Kirchhoff's law $%
U(t)=U_{F}(t)+\Delta U$. The efficiency of a rectifying
device is given by: $\eta =\overline{P}/\overline{P_{F}}=%
-\overline{P}/(\overline{P}+\overline{Q})$. Hence,

\begin{equation}
\eta =-\overline{I(t)}\Delta U/\overline{I(t)U_{F}(t)}.  \label{KPD}
\end{equation}
Note that Eq.(\ref{KPD}) is valid for any one-dimensional rectifying device
where the energy input takes place through the work of the outer forces, the
useful work is produced against the constant field (by pumping particles
uphill) and the Joule heat is dissipated, cf. Refs. \cite{Sok+Blum,%
Takagi1,Takagi2,Mazonka}. We get: 
\begin{equation}
\eta =-\overline{I(\Delta U+U_{F}(t))}\Delta U/\overline{I(\Delta
U+U_{F}(t))U_{F}(t)}  \label{VAC}
\end{equation}
Applying Eq.(\ref{VAC}) to a system rectifying sinusoidal outer field $%
U_{F}(t)=U_{0}\sin \omega t$ one gets after the change of variable $x=\sin
\omega t$: 
\begin{equation}
\eta =-\frac{\int_{-1}^{1}dx\ \xi I\left( U_{0}(x+\xi )\right) /\sqrt{%
1-x^{2}}}{\int_{-1}^{1}dx\ xI\left( U_{0}(x+\xi )\right) /\sqrt{1-x^{2}}},
\label{etasin}
\end{equation}
where $\xi =\Delta U/U_{0}$. In order to understand what property of the
system is important for achieving high efficiencies let us discuss a
hypothetical appliance with a piecewise-linear LCC 
\begin{equation}
I(U)=\left\{ 
\begin{array}{ll}
g_{+}U & {\text{for }} U>0 \\ 
g_{-}U & {\text{for }} U<0
\end{array}
,\right.   \label{pieswise}
\end{equation}
for which Eq.(\ref{etasin}) can easily be evaluated analytically. The
behavior of $\eta $ for $U_{0}=0$ as a function of the outer potential $%
\Delta U$ is shown in Fig.1 for different values of the backward
conductivity $g_{-}$. The larger is the backward resistance, the larger
maximal efficiency is achieved. For $g_{-}\rightarrow 0$ the maximal
efficiency of a gear tends to 1, and is attained for $\Delta U=U_{0}$. In
this case the rectifier is always switched in its backward direction
(locked) so that the stalling case is essentially a no-current one. Thus, if
for an irreversible mode of operation, the stalling condition corresponds to
vanishing of the currents, the losses are suppressed and the ideal
efficiency is reached. This finding can be compared with the results of Ref.%
\cite{Sokolov3}, where the Carnot efficiency is achieved by a heat engine
built of the two ideal diode rectifiers at different temperatures.

Let us now turn to another question: how to build an appliance based on a
ratchet principle, whose efficiency tends to unity under idealized
conditions. We confine ourselves to a quasistatically operating systems as
only candidates for potentially ideal performance. The rocked
one-dimensional ratchet performs badly, because its nonlinearity is too
weak. The LCC of a ratchet rectifier can be obtained by using an adiabatic
solution, Refs. \cite{Magnasco1,Takagi1,Takagi2}, and leads to
practically linear behavior at larger voltages of both signs, showing thus
no locking behavior. As we have seen, locking is important for achieving
high efficiencies. We also note that the phase space of a genuine
ratchet-and-pawl appliance (showing locking) is at least \textit{%
two-dimensional }\cite{Mag+Stol}.

Many typical ratchet appliances, discussed in the literature, can be
considered as special cases of a generic two-dimensional ratchet model with
two impenetrable saw-tooth boundaries in a homogeneous outer field (see
Fig.1a) which we call an oblique rectifier. Standard one-dimensional models
correspond to the case when only a narrow current channel between the
boundaries is present. The system works as a rocked ratchet if only the $x-$%
component of the field oscillates, and as a flashing appliance when only the 
$y$-component changes. The weakness of corresponding nonlinearities is
connected with the fact that the number of the particles in a current
channel does not depend on the field. On the other hand, the system with
broad channel in an oblique field can reach very high efficiencies due to
locking.

Note that an oblique rectifier in homogeneous outer field $\mathbf{F}$ can
be described by a LCC: In homogeneous field one has $Q=\int {\mathbf{jF}}dV%
={\mathbf{F}}\int {\mathbf{j}}dydx$. Due to conservation $\int {\mathbf{j}}dy=I$
so that $Q=I{\mathbf{F}}\int dx=IU$, where $U$ is the potential difference
between the leftmost and the rightmost cross-sections of system. On the
other hand, $P=I\Delta U$ per definition.

In what follows we don't attempt to discuss in detail the LCC of a generic
oblique appliance, and present only a qualitative discussion. In an oblique
field the appliance can effectively be considered as consisting of the
effective current channel and the trapping pockets. Fig.2b) shows a cartoon
corresponding to this strongly simplified picture. The current flowing
through the channel is proportional to the $x$-component of the outer field $%
F$ and to the concentration $n(F)$ of particles in the channel, $I=\mu
SFn(F) $. Here $S$ is the channel's cross-section and $\mu $ is the
particles' mobility. The concentrations of the particles in a channel and at
the opening of the neck connecting it with the pocket are equal; the
particle's concentration in the pocket (having the typical energetic depth $%
\Delta u=-Fd,$ where $d$ is the distance form the neck to the pocket's body)
is $n_{p}(F)=n\exp \left( \Delta u/kT\right) $. Since the overall number of
particles per rectifying unit is field-independent, one has $n(F)\left[
\Omega _{c}+\Omega _{p}\exp \left( Fd/kT\right) \right] =n_{0}(\Omega
_{c}+\Omega _{p})$, where $\Omega _{c}$ and $\Omega _{p}$ are the volumes of
the channel and the pocket, respectively and $n_{0}$ is the
concentration in the absence of the outer field. 
From this the form of the LCC follows as: 
\begin{equation}
I(U)=\frac{g_{0}U}{1+a\exp (-U/U_{T})}.  \label{pocket}
\end{equation}
Here $g_{0}=\mu n_{0}S(1+\Omega _{p}/\Omega _{c})/l$ is the zero-field
conductivity, $U_{T}=kT/d$ is the characteristic field and $a=\Omega
_{p}/\Omega _{c}$. For strong positive fields the behavior of the appliance
is linear. For strong negative fields, the particles get trapped in pockets,
and the current through the appliance decays exponentially. Thus the generic
locking behavior shows up. Numerical evaluation of Eq.(\ref{etasin}) using
the LCC Eq.(\ref{pocket}) leads to results shown in Fig.3. Here we plot $%
\eta (\Delta U)$ as a function of $\Delta U/U_{0}$, for the fixed values of $%
g_{0}=a=U_{T}=1$ and for the values of $U_{0}$ equal to 1, 3, 10 and 30. We
note that the maximal efficiency grows with $U_{0}$ (for $U_{0}=30$ maximal
value of efficiency exceeds 90\%), and the position of this maximum shifts
to the left, i.e. to the values $\Delta U$ approaching $-U_{0}$. The reason
for the growth of efficiency is the fact that typical reverse differential
resistance grows exponentially with $U$, and in the limit of strong outer
fields tends to the ideal limit of 1.

In summary, irreversible gears have considerable advantages when compared
with reversible ones. They allow for an asynchronous mode of operation, of
great virtue for biological systems, since distinct reaction events can
hardly be synchronized on molecular level. We have shown that
irreversibility does not put any limits on the efficiency of energy
transduction, i.e. that an ideal rectifying appliance can reach the
efficiency of 1. The property important for the effective rectification is
locking under backwards load. We show that small modifications of a generic
ratchet system (an oblique rectifying appliance) lead to systems whose
efficiency tends to this idealistic limit.

The author is indebted to Prof. A. Blumen and Prof. J. Klafter for fruitful
discussions. Financial support by the Deutsche Forschungsgemeinschaft
through the SFB 428 and by the Fonds der Chemischen Industrie is gratefully
acknowledged.

\pagebreak

{\Large Figure Captions }

\bigskip

Fig.1. The efficiency of a rectifier
with LCC, Eq(\ref{pieswise}) switched according to a scheme shown as an
insert. The battery is charged when $\Delta U<0$. The fat line corresponds
to an ideal appliance with $g_{-}=0$. Three other curves correspond to $%
g_{-}=10^{-4}$, $10^{-3}$ and $10^{-2}$, respectively. 

\bigskip

Fig.2. a) The oblique rectifying
appliance, see text for details. When the outer field is strong enough the
particles get trapped between the saw-teeth. The trapping potential is
proportional to the outer field $F$. b) The cartoon of the appliance in fig.
a), used in our considerations: this simplified version consists of the
current channel and the pockets, where the particles get trapped if the
outer field shows in reverse direction.

\bigskip

Fig.3. Efficiency of a trapping
rectifier as a function of $\Delta U/U_{0}$ at different values of outer
field amplitude $U_{0}.$ Note that at larger fields the maximal efficiency
tends to unity due to locking.

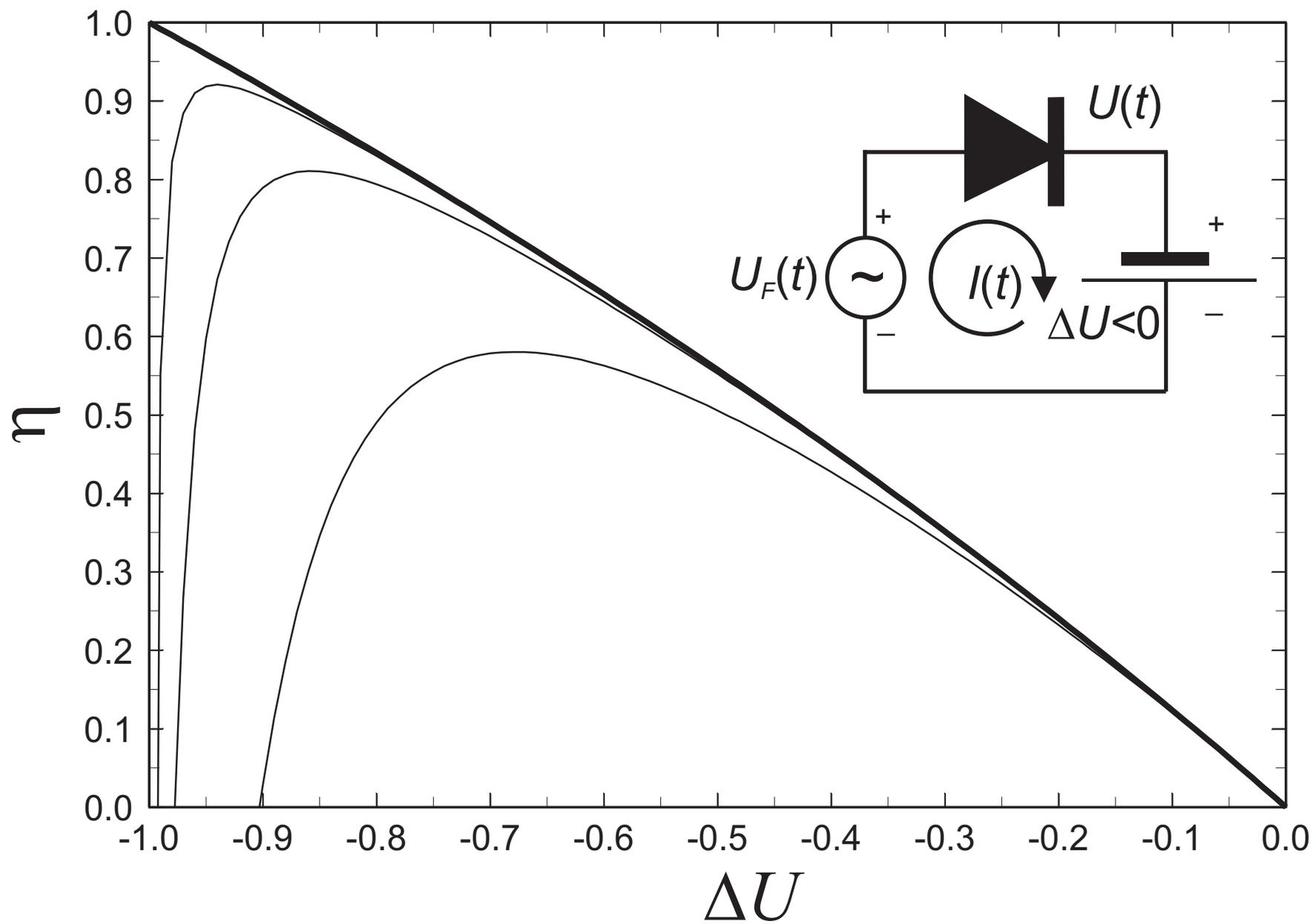
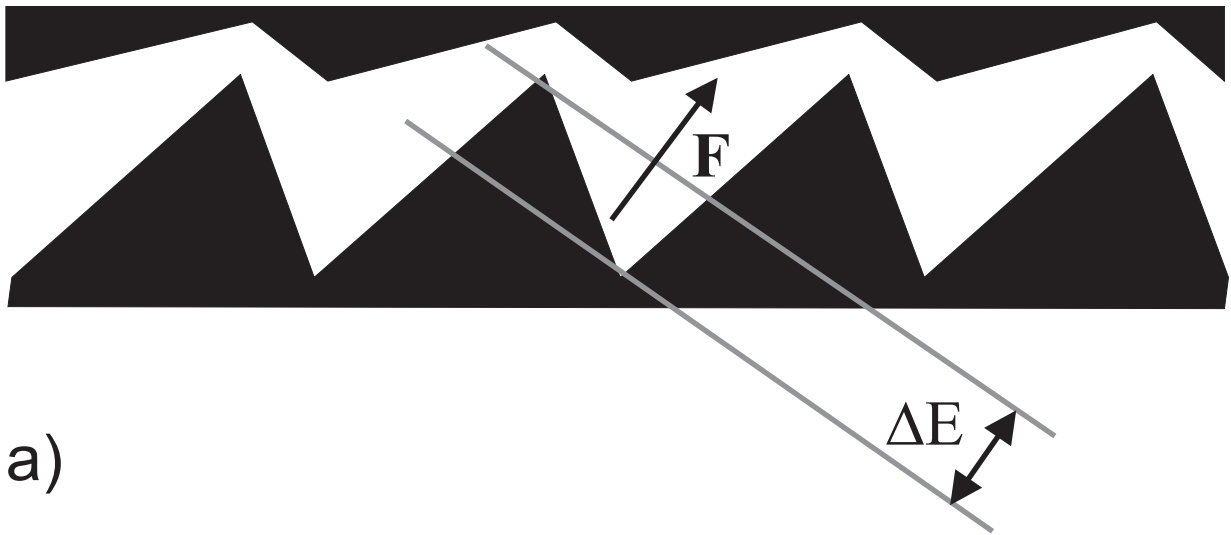
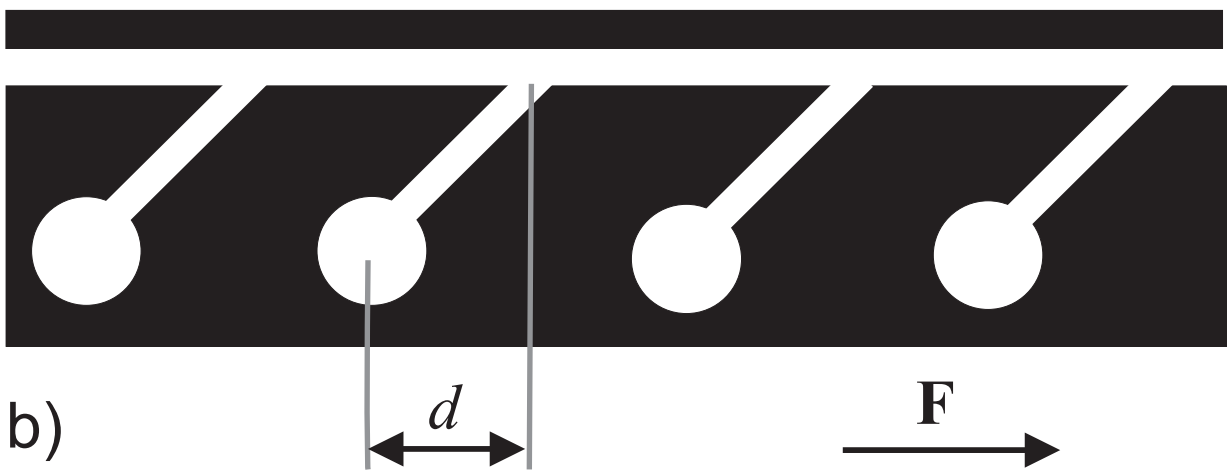
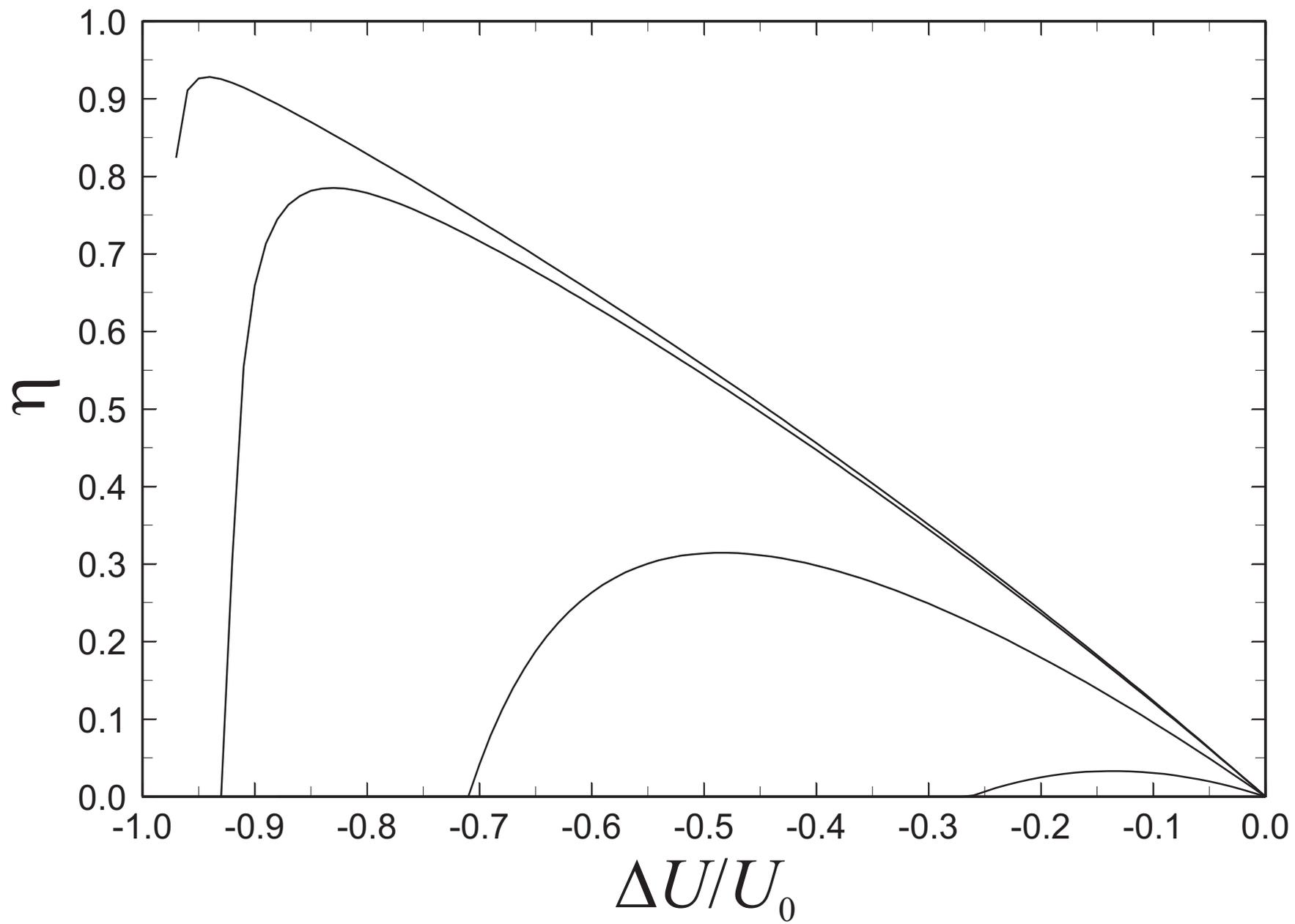
\end{document}